\documentclass[sigconf, screen, 9pt]{acmart}

\usepackage{graphicx}
\usepackage{multirow}
\usepackage{amsmath,amssymb,amsfonts}
\usepackage{amsthm}
\usepackage{mathrsfs}
\usepackage[title]{appendix}
\usepackage{xcolor}
\usepackage{textcomp}
\usepackage{manyfoot}
\usepackage{booktabs}
\usepackage{algorithm}
\usepackage{algorithmicx}
\usepackage{algpseudocode}
\usepackage{listings}
\usepackage{longtable}
\usepackage{lscape}
\usepackage{tabto}
\usepackage{xurl}
\usepackage[nolist,nohyperlinks]{acronym}
\usepackage{booktabs}
\usepackage{adjustbox}
\usepackage{dirtree}
\usepackage{pythonhighlight}
\usepackage{siunitx}
\usepackage{mathtools}
\usepackage{textcomp,mathcomp}

\acrodef{hdd}[HDD]{heating degree day}  
\acrodef{cdd}[CDD]{cooling degree day}  
\acrodef{ami}[AMI]{advanced metering infrastructure}  
\acrodef{ashp}[ASHP]{air-source heat pump}  
\acrodef{cop}[COP]{coefficient of performance}  
\acrodef{dhw}[DHW]{domestic hot water}  
\acrodef{dr}[DR]{demand response}  
\acrodef{eu}[EU]{European Union}  
\acrodef{ev}[EV]{electric vehicle}  
\acrodef{ewh}[EWH]{electric water heater}  
\acrodef{gshp}[GSHP]{ground-source heat pump}  
\acrodef{hp}[HP]{heat pump}  
\acrodef{iea}[IEA]{International Energy Agency}  
\acrodef{ekz}[EKZ]{Elektrizit\"atswerke des Kantons Z\"urich}  
\acrodef{mae}[MAE]{mean absolute error}
\acrodef{mape}[MAPE]{mean absolute percentage error}
\acrodef{mdae}[MdAE]{median absolute error}
\acrodef{mse}[MSE]{mean squared error}
\acrodef{sfoe}[SFOE]{Swiss Federal Office of Energy}
\acrodef{obis}[OBIS]{Object Identification System}
\acrodef{pv}[PV]{photovoltaic}  
\acrodef{sm}[SM]{smart electricity meter} 
\acrodef{smape}[SMAPE]{symmetric mean absolute percentage error}
\acrodef{smd}[SMD]{smart meter data}  
\acrodef{uk}[UK]{United Kingdom}  
\acrodef{utc}[UTC]{coordinated universal time}
\acrodef{us}[US]{United States}

\def\nrhouseholds{1,408}
\def\nrtreatmentgroup{214}
\def\nrtreatmentgroupdoublevisits{3}
\def\nrtreatmentgroupbeforeafter{151}
\def\nrtreatmentgroupbeforeonly{10}
\def\nrtreatmentgroupafteronly{53}
\def\nrcontrolgroup{1,194}
\def\nrweatherstations{8}
\def\nrprotocols{410}

\def\smdfifteenminaveragedays{638.75}

\def\smddailyearliest{2018-11-03}
\def\smddailylatest{2024-03-21}
\def\smddailyaveragedays{721.13}

\begin{document}

\title[HEAPO~--~An Open Dataset for Heat Pump Optimization]{HEAPO~--~An Open Dataset for Heat Pump Optimization with Smart Electricity Meter Data and On-Site Inspection Protocols}


\author{Tobias Brudermueller}
\affiliation{%
  \institution{Chair of Information Management, ETH Zurich}
  \city{Zurich}
  \country{Switzerland}}
\email{tbrudermuell@ethz.ch}
\orcid{0009-0007-7319-1142}

\author{Marina Gonz\'alez Vay\'a}
\affiliation{%
  \institution{New Technology Department, Elektrizit\"atswerke des Kantons Z\"urich}
  \city{Zurich}
  \country{Switzerland}}
\email{marina.gonzalezvaya@ekz.ch}

\author{Elgar Fleisch}
\affiliation{%
  \institution{Chair of Information Management, ETH Zurich}
  \city{Zurich}
  \country{Switzerland}}
\email{efleisch@ethz.ch}
\orcid{0000-0002-4842-1117}
\additionalaffiliation{%
  \institution{Institute of Technology Management, University of St. Gallen}
  \city{St. Gallen}
  \country{Switzerland}
}

\author{Thorsten Staake}
\affiliation{%
  \institution{Chair of Information Systems and Energy Efficient Systems, University of Bamberg}
  \city{Bamberg}
  \country{Germany}}
\email{thorsten.staake@uni-bamberg.de}
\orcid{0000-0003-1399-4676}
\additionalaffiliation{
\institution{Chair of Information Management, ETH Zurich}
  \city{Zurich}
  \country{Switzerland}
}

\renewcommand{\shortauthors}{Brudermueller et al.}

\begin{abstract}

Heat pumps are essential for decarbonizing residential heating but consume substantial electrical energy, impacting operational costs and grid demand. 
Many systems run inefficiently due to planning flaws, operational faults, or misconfigurations.
While optimizing performance requires skilled professionals, labor shortages hinder large-scale interventions. 
However, digital tools and improved data availability create new service opportunities for energy efficiency, predictive maintenance, and demand-side management.
To support research and practical solutions, we present an open-source dataset of electricity consumption from \nrhouseholds{} households with heat pumps and smart electricity meters in the canton of Zurich, Switzerland, recorded at 15-minute and daily resolutions between \mbox{\smddailyearliest{}} and \mbox{\smddailylatest{}}. 
The dataset includes household metadata, weather data from \nrweatherstations{} stations, and ground truth data from \nrprotocols{} field visit protocols collected by energy consultants during system optimizations.
Additionally, the dataset includes a Python-based data loader to facilitate seamless data processing and exploration.


\end{abstract}



\settopmatter{printacmref=false}
\setcopyright{none}
\renewcommand\footnotetextcopyrightpermission[1]{}
\pagestyle{plain}

\maketitle

\section*{General Note}
Please note that this manuscript on arXiv is a preprint.  
The dataset and dataloader are already available in their initial version, but updates may occur in future releases as the manuscript is currently under peer review.  
If you use the dataset in its initial form, please cite this arXiv paper.
\section{Introduction}
\label{sec:introduction}
Electric \acp{hp} are a key solution for decarbonizing residential space and water heating.
However, their complexity compared to traditional gas or oil-based systems makes real-world efficiency heavily reliant on proper planning \cite{DONGELLINI2021114022, RENALDI2017520}, optimal settings \cite{tejeda2014energy}, and fault-free operation \cite{MADANI201419, HU2021110975}, which many users and installers find difficult to manage \cite{DECUYPERE2022112709, caird2012domestic}.
Consequently, real-world performance varies widely, and many \acp{hp} have substantial room for optimization.
For example, \citet{NOLTING2018476} found real-world efficiency to be up to 24\% lower than rated values.
Likewise, a study of 297 Swiss households showed that after optimization, half reduced annual electricity use by 1,805 kWh (15.2\%) on average \cite{weigert2022heat}.
Unexploited optimization potential from \acp{hp} can strain electricity grids \cite{10407931, LIZANA2023125298, LOVE2017332, rinaldi2022adds} and drive up homeowners' costs, undermining economic viability and hindering the transition to sustainable heating \cite{engelen2023heat, nast2007instruments, bauermann2016german, decker2015house}. 
In this context, digital tools and data-driven solutions provide valuable opportunities for predictive maintenance, enhanced energy efficiency, and demand-side flexibility to address these challenges. 
Studies show that \acp{hp} can even enhance grid stability, with operational flexibility supporting frequency regulation and reserve capacity in demand response programs \cite{BAUMANN2023100081, 10.1115/1.4045819, KRUG2025101662}. 
For example, throttling 300+ \acp{hp} reduced load by 40-65\% \cite{muller2019large}, and flexible \acp{hp} in Germany lowered homeowners' electricity costs by 4-9\% in \acl{dr} scenarios \cite{GLOBISCH2020101604}.

Digital solutions for \ac{hp} optimization are even more important in the light of a shortage of skilled workers. 
The \acl{eu} estimates a need for 750,000 additional installers, with at least 50\% of the current workforce requiring reskilling \cite{eu_heatpumps}. 
In the \acs{uk}, where \acp{hp} can only be installed by certified professionals, only 1,800 installers were certified in 2020, far below the 69,500 needed by 2035 \cite{060f1038ba8c49dbb129f054a351e168}. 
This labor shortage affects both the deployment of new \acp{hp} and the optimization of existing systems. 
To address this, digital tools are essential for assessing system performance, identifying households with optimization potential, and prioritizing them for on-site interventions by professionals. 
While many modern \acp{hp} already transmit high-quality sensor data, including pressure, volume flow, and energy input/output, a significant number of existing units lack internet connectivity and advanced measurement capabilities \cite{fischer2017business, ieaannex56}.
Yet, they will remain in operation for decades. 
In such cases, \ac{ami} with \acp{sm} provides a scalable alternative for remote \ac{hp} monitoring \cite{pub:hptype, pub:hpdisaggregation, muller2019large}. 
For example, a study using real-world \ac{smd} from 503 Swiss households has already demonstrated that inefficient \acp{hp} can be identified by analyzing their on-off cycling behavior \cite{pub:hpcycling}.

To drive research advancements and scale real-world services, which still remain rarely deployed, high-quality data must be freely available to capture the variability of real-world system performance \cite{miller2020building, articlea:hopf2018enhancing, articled:hopf2022value}.
Understanding the factors that influence \ac{hp} behavior and performance, as well as training algorithms to capture diverse usage patterns, requires a dataset that combines operational data with contextual information about the building and heating system, along with reliable ground-truth data on \ac{hp} settings and optimization potential.
There are several open-source datasets related to residential electricity consumption and \ac{hp} usage, but none provide all the functionalities required for developing algorithms tailored to the aforementioned purposes \cite{schlemminger2022dataset, articlea:ruhnau2019time, agee2021measured, pereira2022residential, klemenjak2020synthetic, quesada2024electricity}.  
For instance, the dataset by \citet{schlemminger2022dataset} offers residential electricity and \ac{hp} load profiles from single-family homes in Northern Germany, recorded at resolutions ranging from 10 seconds to 60 minutes between 2018 and 2020. 
However, with data from only 38 households, the sample size is small, and the dataset includes very limited metadata on household characteristics, with no details on \ac{hp} settings.  
Another well-known dataset in the \ac{hp} domain is \emph{When2Heat} by \citet{articlea:ruhnau2019time}, which provides synthetic national time series for heat demand and \ac{hp} coefficient of performance across 16 European countries rather than real-world measurements. 
Other commonly used \ac{smd} datasets, often applied in non-intrusive load monitoring research, include \emph{REDD} \cite{articlec:kolter2011redd}, \emph{UK-DALE} \cite{articlec:kelly2015uk}, \emph{GREEND} \cite{articlec:monacchi2014greend}, and \emph{AMPds} \cite{articlec:makonin2016electricity}. 
Among these, only \emph{AMPds} contains \ac{hp} measurements, providing one-minute power readings for a single house in Canada over two years.  
In summary, there is currently no publicly and freely available dataset with a large sample size that integrates \ac{smd}, metadata, and ground-truth data from residential buildings using \acp{hp} for heating or cooling.

To address these challenges, this work presents \emph{HEAPO (Heat Pump Optimization)} -- a real-world, open-source dataset designed to support researchers, practitioners, utilities, and other stakeholders in advancing both research and practical solutions for optimizing \ac{hp} operations.  
For instance, \emph{HEAPO} can be utilized to develop advanced algorithms, applications, and services for improving energy efficiency, disaggregating loads, leveraging flexibility in demand response programs, and estimating residential \ac{pv} feed-in to the grid.
The dataset provides cleaned and ready-to-use energy data at both 15-minute and daily resolutions, recorded by standard smart electricity meters in \nrhouseholds{} real households that use \acp{hp} for space and water heating.
These households are located in the canton of Zurich, Switzerland. 
The field data was collected between \smddailyearliest{} and \smddailylatest{}, with each household having an average of \smddailyaveragedays{} days of available data in daily resolution and \smdfifteenminaveragedays{} days in 15-minute resolution.
Additionally, the dataset includes temperature data from the \nrweatherstations{} nearest weather stations matched to household locations, metadata provided by households through an online tool from their utility company, and \nrprotocols{} field visit protocols recorded by professional energy consultants.
These consultants, with over a decade of experience conducting energy audits, optimized the \acp{hp} during their on-site visits. 
For \nrtreatmentgroup{} of the households with available \ac{smd}, a corresponding audit protocol is available. 
Therefore, the protocols provide valuable insights into common issues observed in real-world installations and enable the analysis of changes in electricity consumption following the optimization, offering a unique opportunity to evaluate the impact of targeted interventions.
To facilitate seamless use of the dataset, a Python-based dataloader including visualization functionalities is provided, enabling efficient data processing and software development without delay. 
Both the dataset and the accompanying code are publicly available on GitHub: \href{https://github.com/tbrumue/heapo}{https://github.com/tbrumue/heapo}
\section{Dataset Description}
\label{sec:methods}

This section outlines data collection and pre-processing, dataset structure, dataloader functionality, and technical validation.

\subsection{Data Collection and Pre-Processing}
\begin{figure}[ht!]
\centering 
\includegraphics[width=\columnwidth]{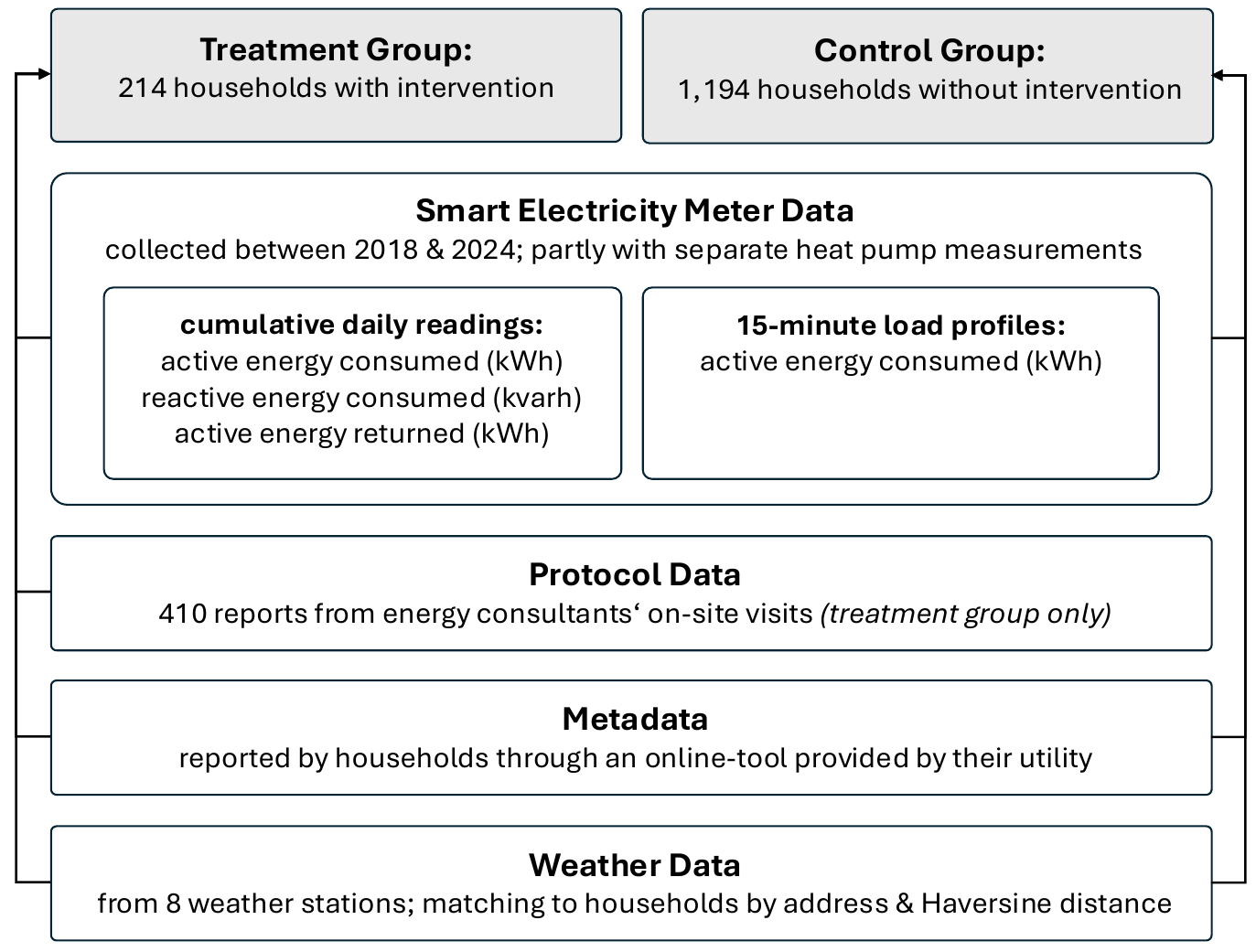}
\caption[]{Overview of available data sources.} 
\label{fig:overview}
\end{figure}

The dataset integrates information from multiple sources into a user-friendly structure, assigning each household a unique identifier (\emph{Household\_ID}) for seamless cross-referencing across all data instances.  
To ensure compliance with data privacy regulations, identifiers are randomly assigned integer values, and any information that could reveal a household's exact location or private details has been removed.
Details on data collection and pre-processing steps for each source are provided in the following subsections, with an overview illustrated in Figure~\ref{fig:overview}.

\subsubsection{Protocol Data}
The dataset includes \nrprotocols{} reports from on-site energy audits conducted between 2015 and 2024 by experienced energy consultants specializing in \ac{hp} optimization. 
These audits are part of a paid service provided by a local energy consulting company within the utility and can take place on any day of the year. 
During visits, consultants assess the \ac{hp} and heat distribution system, addressing minor issues that do not require an installer. 
If basic settings such as the heating curve, time programs, or \acl{dhw} temperature are found to be suboptimal, they may adjust them with customer consent. 
However, consultants do not perform retrofits, though they may recommend them. Since customers can adjust settings before or after the visit, the configuration details reflect the system's status at the time of inspection, while metadata about the building typically remains stable. 
We emphasize that the provided reports make this dataset unique, offering valuable insights into common issues observed in field-installed \acp{hp} and providing high-quality ground-truth data to develop algorithms that can anticipate these issues and assess \ac{hp} performance in operation.
Each protocol is assigned a unique identifier, referred to as \emph{Report\_ID}.
For \nrtreatmentgroup{} households, both protocol data and \ac{smd} are available, forming the treatment group, while the remaining \nrcontrolgroup{} households with only \ac{smd} serve as the control group. 
Note that the protocol data also includes 196 reports from households without associated \ac{smd}, which are not part of either the treatment or control group.
For each protocol associated with a treatment group household, a mapping between the report identifier and the household identifier is provided.
Within the treatment group, \nrtreatmentgroupdoublevisits{} households booked the \ac{hp} service multiple times, resulting in reports from different periods. %
Additionally, \nrtreatmentgroupbeforeonly{} households have \ac{smd} only before the visit, \nrtreatmentgroupafteronly{} only after, and \nrtreatmentgroupbeforeafter{} for both before and after the visit.
In particular, the latter enable a comparison of consumption patterns before and after intervention, allowing for the analysis of associated treatment effects.
A detailed list of variables included in the protocols can be found in Table~\ref{table:reportsvariablesdescription} in the~\nameref{sec:appendix}.
The reports are provided in a machine-readable format, using categorical, binary, or numerical values.  
To preserve household privacy, any commentary data in German has been excluded.

\subsubsection{Smart Meter Data}
Each \ac{sm} in the study's households transmits recorded measurements to the utility company on a daily basis.  
Consequently, missing or corrupted data is limited to full daily records and may appear as continuous periods of zero values or negative readings for grid consumption.  
Such erroneous measurements have been excluded to ensure a cleaned dataset. 
All timestamps in the dataset files are reported in \ac{utc}.  
To facilitate data filtering, each \ac{smd} file includes a \emph{Group} column, indicating whether a household belongs to the treatment or control group, and an \emph{AffectsTimePoint} column, specifying whether each timestamp falls before, during, or after an energy consultant's visit for treatment group households, or is marked as unknown for control group households.
The \acp{sm} provide cumulative daily meter readings since installation, ensuring high data quality for billing by reconciling any missing interval data in subsequent readings.
These cumulative daily readings capture a household's interaction with the power grid, including both energy drawn from the grid and self-generated energy fed back into it in homes with a \ac{pv} system. 
However, they do not capture self-consumed \ac{pv} energy that does not reach the grid.
For the cumulative daily values of received energy, the meters record both active consumption (in kWh) and reactive consumption (in kvarh), distinguishing between inductive (first quadrant) and capacitive (fourth quadrant) components. 
The dataset includes the original cumulative readings along with pre-processed files for daily and monthly consumption.  
Daily consumption is calculated by subtracting the previous day's counter value, while monthly consumption is derived from the difference between the last recorded values of each month, where available.
In addition to the daily cumulative readings for billing purposes, the \acp{sm} also provide 15-minute load profiles of active energy consumption (in kWh), which are calculated by multiplying the average active power (in kW) over each 15-minute interval by the time duration. 
This calculation is performed directly on the device by each \ac{sm} before transmission.
Note that energy returned to the grid and reactive energy received from the grid are only available in the daily cumulative readings, not in the 15-minute load profiles. 
Additionally, some households have two \acp{sm}, one exclusively measuring \ac{hp} electricity consumption and the other covering all other appliances. 
In such cases, data from both meters are included separately, with their sum at each timestamp representing total consumption.
Table~\ref{table:smdvariables} summarizes the pre-processed \ac{sm} measurements provided in the dataset files, along with the corresponding \acs{obis} codes \cite{obiscode, 10.1007/978-981-10-8249-8_2}, the standard used by utilities in German-speaking countries to categorize different types of measurements (see \nameref{sec:appendix:obiscodes} in the \nameref{sec:appendix}).

\begin{table*}[ht!]
\centering
\caption{Description of \acl{smd} variables. Active electrical energy received from the grid (in kWh) is available in both 15-minute and daily resolution, while active electrical energy returned to the grid (in kWh) and reactive electrical energy received (in kvarh) are only available in daily resolution.}
\label{table:smdvariables}
\resizebox{\textwidth}{!}{%
\begin{tabular}{@{}lll@{}}
\toprule
\textbf{\acs{smd} Variable Name} & \textbf{Variable Description} & \textbf{OBIS Code} \\ 
\midrule
Household\_ID & unique household identifier & \\ 
AffectsTimePoint & timestamp category relative to energy consultants' visit date (before, after, during, or unknown) & \\ 
Timestamp & \acs{utc} timestamp of data measurement (\textit{YYYY-MM-DD HH:MM:SS}) & \\ 
kWh\_received\_Total & household's total active energy consumed & 1-1:1.8.0 / 1-1:1.29.0 \\ 
kWh\_received\_HeatPump & when separate \acs{sm}: active energy consumed by \acs{hp} & 1-1:1.8.0 / 1-1:1.29.0 \\ 
kWh\_received\_Other & when separate \acs{sm}: active energy consumed by appliances other than \ac{hp} & 1-1:1.8.0 / 1-1:1.29.0 \\ 
kWh\_returned\_Total & household's total active energy returned (if \ac{pv} system available) & 1-1:2.8.0 \\ 
kvarh\_received\_capacitive\_Total & household's total reactive capacitive energy consumed & 1-1:8.8.0 \\ 
kvarh\_received\_capacitive\_HeatPump & when separate \acs{sm}: reactive capacitive energy consumed by \acs{hp} & 1-1:8.8.0 \\ 
kvarh\_received\_capacitive\_Other & when separate \acs{sm}: reactive capacitive energy consumed by appliances other than \ac{hp} & 1-1:8.8.0 \\ 
kvarh\_received\_inductive\_Total & household's total reactive inductive energy consumed & 1-1:5.8.0 \\ 
kvarh\_received\_inductive\_HeatPump & when separate \acs{sm}: reactive inductive energy consumed by \acs{hp} & 1-1:5.8.0 \\ 
kvarh\_received\_inductive\_Other & when separate \acs{sm}: reactive inductive energy consumed by appliances other than \ac{hp} & 1-1:5.8.0 \\ 
\bottomrule
\end{tabular}%
}
\end{table*}

\subsubsection{Weather Data}
Heating and cooling activity often correlates with outdoor temperature conditions \cite{WESTERMANN2020114715}.  
To account for this temperature dependence in \ac{smd} analysis, the dataset includes weather data from \nrweatherstations{} nearby weather stations.  
Using each household's address, we matched it to the nearest weather station based on haversine distance and provided a mapping between household identifiers and their corresponding weather station identifiers.
The weather data is sourced from MeteoSwiss \cite{meteoschweiz}, a highly reputable nationwide weather service, where data is available for download upon registration.
It includes daily and hourly measurements of key meteorological variables, such as maximum, minimum, and average temperatures, humidity, precipitation, sunshine duration, wind speed, pressure, and dew point temperature.
Table~\ref{table:weathervariables} in the \nameref{sec:appendix} presents a comprehensive description of the available weather data variables.
Note that MeteoSwiss derives \ac{hdd} and \ac{cdd} numbers from the average daily temperature ($T_\text{avg}$). 
The estimated heating energy demand for a building at a given location is assumed to be directly proportional to its \ac{hdd} value. 
MeteoSwiss follows standards defined for both the United States and Switzerland (SIA 381/3 \cite{articlea:standard1982381}), as given by the formulas below.

\begin{equation}
\label{formula:hddsia}
\text{\acs{hdd} (Switzerland)} = 
    \begin{dcases*}
    20~\text{\textdegree C} - T_\text{avg} & \text{if } $T_\text{avg} > 12$ \\
    0 & \text{otherwise}
    \end{dcases*} 
\end{equation}

\begin{equation}
\label{formula:hddus}
\text{\acs{hdd} (United States)} = \text{max}(0, 18.3~\text{\textdegree C} - T_\text{avg})
\end{equation}

\begin{equation}
\label{formula:cddus}
\text{\acs{cdd} (United States)} = \text{max}(0, T_\text{avg} - 18.3~\text{\textdegree C})
\end{equation}


\subsubsection{Metadata} 
\label{sec:metadata}
All households can access their electricity consumption data from \acp{sm} free of charge via an online customer portal provided by their utility company.  
Within this portal, they can voluntarily enter metadata to receive personalized energy efficiency recommendations.  
Examples of collected metadata variables include the number of residents and the type of heat distribution system. 
A complete list of available variables is provided in Table~\ref{table:metadatavariables} in the \nameref{sec:appendix}.  
The metadata is available for all households, regardless of group assignment.
However, it is included in the \emph{HEAPO} dataset to offer contextual insights that enhance advanced \ac{smd} analytics, particularly for households in the control group where protocol data is not available.  

\subsection{Dataset Structure}
\label{sec:datasetstructure}
The dataset is organized in an intuitive folder and file structure (see illustration in Figure~\ref{fig:filestructure} in the \nameref{sec:appendix}) to minimize barriers to data exploration.  
This structure allows users to quickly grasp the context without needing to reference this paper in detail, making it especially useful for those working in programming languages other than Python, for which we provide code for data loading and exploration (see the following section). 
The data is divided into four main subfolders, corresponding to different sources: \emph{meta\_data}, \emph{reports}, \emph{smart\_meter\_data}, and \emph{weather\_data}.  
Each subfolder is accompanied by descriptive files that clarify data availability and explain the meaning of each column or variable.  
Metadata and protocol data are provided in single CSV files for ease of access.  
In contrast, weather data is split into separate CSV files per weather station, organized into subfolders based on temporal resolution (daily and hourly).  
Similarly, each household's \ac{smd} is stored in a separate CSV file, grouped into subfolders by resolution (15-minute, cumulative counters, daily, and monthly).

\subsection{Python-Based Dataloader}
\label{sec:methods:dataloader}
\begin{figure*}[ht!]
\centering 
\includegraphics[width=\textwidth]{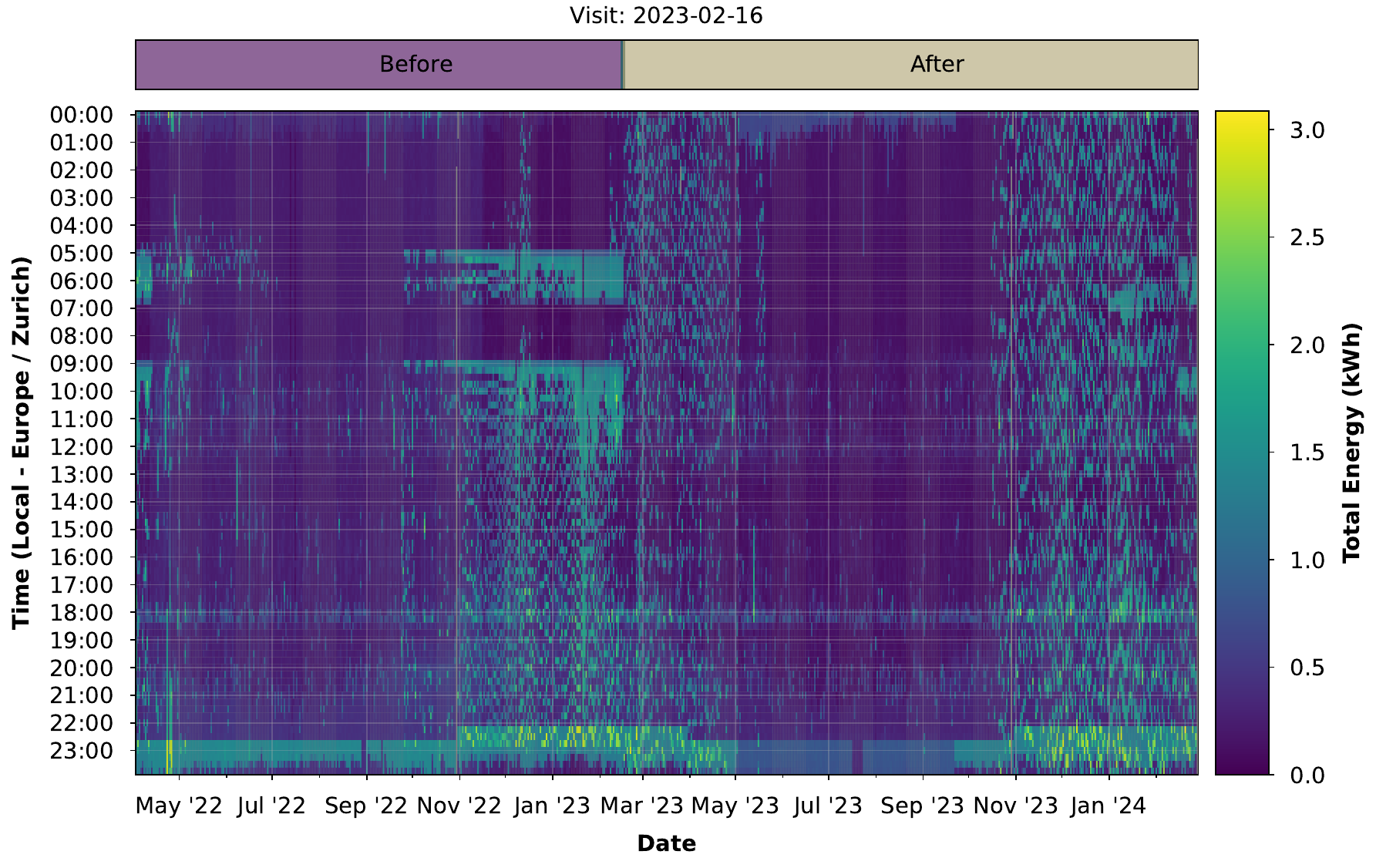}
\caption[]{Example of 15-minute resolution \ac{smd} for a single household (Household\_ID: 461158) visualized as a heat map. The graph shows total active electrical energy consumption (in kWh) using colored pixels, with the date on the x-axis and the time of day on the y-axis. A shift in consumption patterns is noticeable after the energy consultant's visit, where the night setback was deactivated as part of the \ac{hp} optimization.} 
\label{fig:heatmap_example}
\end{figure*}

The dataset is accompanied by a Python script that provides a data loading class with built-in functionalities for intuitive data exploration, visualization, and filtering.
For example, a single function call can load \ac{smd} and weather data into a unified pandas DataFrame, where weather data is automatically interpolated from hourly to 15-minute resolution if the \ac{smd} has the latter resolution.
Additionally, time series data for treatment households can be easily segmented into periods before, during, and after an energy consultant's on-site visit or filtered to include only a specific number of months before and after consultation.
Another function enables adjusting the temporal resolution of the data, allowing users to aggregate \ac{smd} into hourly or weekly intervals, even if these were not originally measured.
Figure~\ref{fig:heatmap_example} showcases a heatmap visualization of a treatment household's \ac{smd} before and after consultation, generated using our script. 
The x-axis represents dates, the y-axis represents time of day, and energy consumption (in kWh) is color-encoded as pixels.
With the provided Python script, generating such a figure requires only a few lines of code, as demonstrated in Figure~\ref{fig:pythonexample}.

\tiny
\begin{figure*}[ht!]
\begin{minipage}{\textwidth}
\begin{python}
# import code and load data
from heapo import * 
heapo=HEAPO(data_path='/path/to/datafolder/', use_local_time=False) 
household_id=461158 # define household of interest
df=heapo.load_smart_meter_data(household_id,resolution='15min') 
report_date=heapo.load_protocol_data(household_id)['Visit_Date'].values[0]

# create figure for heatmap, colorbar and data range
fig,ax=plt.subplots(2,2,figsize=(12,8),gridspec_kw={'height_ratios':[0.3,4],'width_ratios':[1,0.05]})

# plot heatmap, color bar, and consultation range
hm = heapo.plot_heatmap(df,'kWh_received_Total',ax=ax[1,0],hour_interval=1,fontsize=14)
heapo.plot_cbar_on_ax(hm['mesh'],cax=ax[1,1],fontsize=14,cbarlabel='Total Energy (kWh)')
heapo.plot_consultation_range_on_ax(ax[0,0],df,report_date,fontsize=14)

# finalize figure formatting
ax[0,1].set_visible(False)
ax[0,0].set_title('Visit: {}'.format(report_date),fontsize=14,pad=10) 
fig.suptitle('Household: {}'.format(household_id),fontsize=16,fontweight='bold') 
fig.subplots_adjust(top=1.2)
fig.tight_layout()
plt.show()
\end{python}
\end{minipage}
\caption{Example of Python code demonstrating the use of the dataloader and its data visualization features. This code snippet produces the graph shown in Figure~\ref{fig:heatmap_example}.}
\label{fig:pythonexample}
\end{figure*}
\normalsize

\subsection{Technical Validation}
For the technical validation of the \ac{smd}, we assess the consistency of total active energy consumption measurements (kWh\_received\_Total) between daily data and 15-minute load profiles. 
First, we aggregate the 15-minute load profiles to a daily resolution by summing the values for each household and date. 
Next, we match the measurements by household and date, retaining only observations where at least 10 days of overlapping data exist between the summed 15-minute data and the daily data. 
This results in 437,592 days of observations from 866 households included in the evaluation. 
To account for variations in data availability and potential noise levels among different \acp{sm}, we compute error levels individually for each household, reporting the mean and standard deviation across all households. 
The errors are calculated by treating one dataset as the ground truth and the other as the predicted values.
Considering that daily cumulative sums are used for billing, they can be regarded as high-quality data. 
The validation yields an $R^2$ score of $0.99 \pm 0.01$ per household and day, a \acl{mae} of $0.48 \pm 0.12$ kWh, and a \acl{smape} of $2.46 \pm 0.69~\%$. 
Additionally, we validate the weather data by analyzing variations in daily average temperature (\emph{Temperature\_avg\_daily}) across different weather stations. 
We compute the standard deviation of daily temperature measurements among the eight stations and evaluate 1,844 days where data is available for all stations. 
The analysis reveals an average inter-station difference of $1.06 \pm 0.31 \text{\textdegree}C$.
The complete implementation details and code for reproducing these results are available in the GitHub repository of the data loader.
\section{Descriptive Analyses}
\label{sec:descriptiveanalyses}

This section provides an overview of the study population, analyzes household energy consumption, and highlights common \ac{hp} issues identified during on-site visits.

\subsection{Study Population Statistics}
Of all households in the dataset, 1,358 (96.45\%) have submitted some \nameref{sec:metadata} through their utility's online tool.
Table~\ref{table:metadataoverview} summarizes key statistics for all households, as well as for the treatment and control groups separately.  
Approximately 55.26\% use an air-source \ac{hp}, 40.91\% rely on a ground-source \ac{hp}, while the remaining households have unspecified \ac{hp} types.  
The majority of households (60.44\%) use their \ac{hp} also for \ac{dhw} production, while 30.11\% rely on an \ac{ewh}.
Additionally, 24.36\% have indicated to own at least one \ac{ev}, and 34.65\% have a \ac{pv} system, as inferred from \ac{smd} measurements indicating energy returned to the grid.
The studied households have an average of 2.87 residents (standard deviation:~1.1) and a mean living area of 193.28~$m^2$ (standard deviation:~71.65).

\begin{table}[ht]
\centering
\caption{Overview of population based on collected metadata.}
\label{table:metadataoverview}
\resizebox{\columnwidth}{!}{%
\begin{tabular}{lccc}
\hline
\multicolumn{1}{c}{\textbf{Description}} & \textbf{\begin{tabular}[c]{@{}c@{}}All\\ Households\end{tabular}} & \textbf{\begin{tabular}[c]{@{}c@{}}Treatment\\ Group\end{tabular}} & \textbf{\begin{tabular}[c]{@{}c@{}}Control\\ Group\end{tabular}} \\ \hline
Households - Total & 1,408 & 214 & 1194 \\
Households - With 15-min \acs{smd} & 1,407 (99.93 \%) & 214 (100.0 \%) & 1,193 (99.92 \%) \\
Households - With daily \acs{smd} & 1,298 (92.19 \%) & 156 (72.9 \%) & 1,142 (95.64 \%) \\
Households - With air-source \acs{hp} & 778 (55.26 \%) & 109 (50.93 \%) & 669 (56.03 \%) \\
Households - With ground-source \acs{hp} & 576 (40.91 \%) & 52 (24.3 \%) & 524 (43.89 \%) \\
Households - With \acs{pv} system & 516 (36.65 \%) & 49 (22.9 \%) & 467 (39.11 \%) \\
Households - With \acs{ev} & 343 (24.36 \%) & 33 (15.42 \%) & 310 (25.96 \%) \\
Households - With dryer & 1081 (76.78 \%) & 119 (55.61 \%) & 962 (80.57 \%) \\
Households - With freezer & 1233 (87.57 \%) & 139 (64.95 \%) & 1094 (91.62 \%) \\
Households - With radiators & 394 (27.98 \%) & 48 (22.43 \%) & 346 (28.98 \%) \\
Households - With floor heating & 1062 (75.43 \%) & 117 (54.67 \%) & 945 (79.15 \%) \\
Households - With \acs{dhw} by \acs{hp} & 851 (60.44 \%) & 112 (52.34 \%) & 739 (61.89 \%) \\
Households - With \acs{dhw} by \acs{ewh} & 424 (30.11 \%) & 44 (20.56 \%) & 380 (31.83 \%) \\
Households - With \acs{dhw} by solar & 70 (4.97 \%) & 7 (3.27 \%) & 63 (5.28 \%) \\
Number of residents (Mean $\pm$ Std) & $2.87 \pm 1.1$ & $2.78 \pm 1.11$ & $2.88 \pm 1.1$ \\
Living area in $m^2$ (Mean $\pm$ Std) & $193.28 \pm 71.65$ & $187.27 \pm 60.39$ & $194.11 \pm 73.04$ \\ \hline
\end{tabular}%
}
\end{table}

\subsection{Household Electricity Consumption}
\begin{figure}[ht!]
\centering 
\includegraphics[width=\columnwidth]{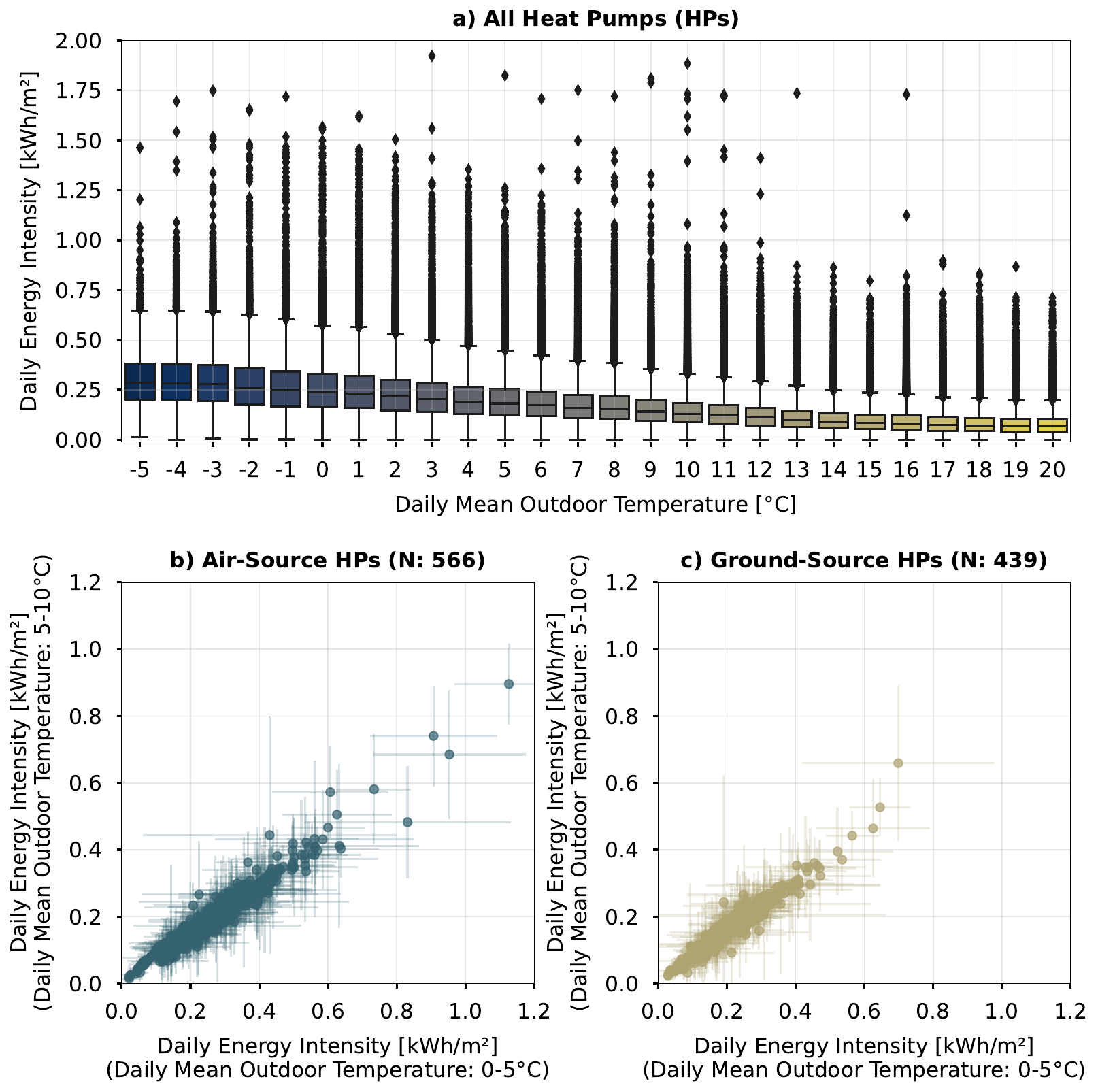}
\caption[]{Daily household energy intensity (kWh per square meter of living area) as a function of outdoor temperature for households in the dataset.} 
\label{fig:energyconsumption}
\end{figure}

Daily household electricity consumption in the \emph{HEAPO} dataset is influenced by outdoor temperature, as heat demand supplied by the \acp{hp} varies with ambient conditions.
Additionally, factors such as building insulation, number of residents, living area, and personal habits further impact energy use \cite{articleb:beckel_revealing_2014}.  
While the metadata accompanying the \ac{smd} enables a detailed analysis of these factors, this paper focuses solely on introducing the dataset and does not explore them in depth. 
However, to provide an overview of electricity consumption in this dataset, Figure~\ref{fig:energyconsumption} presents different visualizations of daily household energy intensity (kWh per square meter), calculated as the total active electrical energy consumed divided by the reported living area.  
This metric is used to normalize electricity consumption relative to building size.
Figure~\ref{fig:energyconsumption}a displays boxplots of daily energy intensity across all households as a function of mean outdoor temperature, rounded to integer values. 
In contrast, Figures~\ref{fig:energyconsumption}b and \ref{fig:energyconsumption}c differentiate by \ac{hp} type, showing average energy intensity per household for temperature ranges of 0-5\textdegree C and 5-10\textdegree C. 
Only households with a minimum of 10 days of data in each temperature range are included.
Each scatter point represents a household, with horizontal and vertical error bars indicating individual standard deviations.  
All graphs reveal a largely linear relationship between daily energy intensity and outdoor temperature, consistent with previous findings in \cite{pub:hpcycling}. 
Additionally, the lower two graphs reveal significant variations in individual household consumption, indicating that factors beyond building size should be considered when developing algorithms to assess \ac{hp} energy efficiency using \ac{smd}.

\subsection{Common Issues Identified in Protocols}
Building on the analysis by \citet{weigert2022identification}, which categorizes common field issues with \acp{hp}, we summarize the most frequently reported problems from energy consultants' on-site visits in Table~\ref{table:commonproblems}.  
The absolute and relative frequencies are based on the \nrprotocols{} reports included in the \emph{HEAPO} dataset.  
Several of the most common issues directly impact energy efficiency. 
The heating curve, which determines the relationship between outdoor temperature and the flow temperature of the heating system, is often set too high. 
This results in excessive energy consumption, as the \ac{hp} generates more heat than necessary. 
Additionally, night setback is frequently activated, causing floor and wall surfaces to cool overnight and requiring extra energy for reheating in the morning. 
Another common issue is that the heating limit setting, which defines the outdoor temperature below which heating is activated, is often set too high, leading to unnecessary heating.  
These findings highlight the need for algorithms that assess \ac{hp} settings for optimal efficiency. 
Moreover, educating users and installers on proper configuration is essential, as familiarity with the technology has been shown to translate to better \ac{hp} performance \cite{caird2012domestic, 10.1145/2674061.2674067, articlea:doi:10.1177/0162243920978301, gleeson2017analysis}.

\begin{table}[ht]
\centering
\caption{Common issues identified by energy consultants in protocol data from on-site visits.}
\label{table:commonproblems}
\resizebox{\columnwidth}{!}{%
\begin{tabular}{lcc}
\hline


\multicolumn{1}{c}{\textbf{Reported Issue}} & \textbf{\begin{tabular}[c]{@{}c@{}}Relative\\ Frequency\end{tabular}} & \textbf{\begin{tabular}[c]{@{}c@{}}Absolute\\ Frequency\end{tabular}} \\ \hline
Heating curve setting chosen too high & 40,98 \% & 168 \\
Night setback mistakenly activated & 36,10 \% & 148 \\
Heating limit setting chosen too high & 25,61 \% & 105 \\
Descaling of \acs{dhw} necessary & 17,80 \% & 73 \\
Expansion system configured inappropriately & 13,41 \% & 55 \\
Problems with air ducting of air-source \acsp{hp} & 11,25 \% & 27 \\
\acs{hp} inappropriately sized & 10,00 \% & 41 \\
Pipes not insulated well enough & 9,02 \% & 37 \\
\acs{dhw} temperature misconfigured & 7,80 \% & 32 \\
Brine pressure not alright for ground-source \acsp{hp} & 7,19 \% & 12 \\
Circulation pump regulation set incorrectly & 6,34 \% & 26 \\
Geothermal probe temperature too low for ground-source \acsp{hp} & 5,99 \% & 10 \\
Install thermostatic valves for individual room regulation & 4,88 \% & 20 \\
\acs{hp} is dirty & 3,17 \% & 13 \\
\acs{hp} technically not in order & 1,46 \% & 6 \\
Basic functions of \acs{hp} are not in order & 1,46 \% & 6 \\ \hline
\end{tabular}%
}
\end{table}
\section{Conclusion}
\label{sec:discussionandconclusion}
The \emph{HEAPO} dataset provides real-world measurements of energy consumption and grid feedback, collected from \nrhouseholds{} households in the canton of Zurich, Switzerland, between 2018 and 2024. 
All households utilize \acp{hp} for space and water heating. 
In addition to \acl{smd} data, the dataset includes metadata, weather data, and protocol data from on-site visits, where energy consultants optimized \ac{hp} systems.  
To support research and analysis, the dataset is accompanied by a Python-based dataloader that facilitates data exploration. 
While the dataset's geographic focus on Switzerland may limit its direct applicability to regions with different climates and building standards, we believe it provides a valuable foundation for advancing \ac{hp} optimization.  
Potential applications include estimating energy savings pre-intervention, assessing achieved treatment effects post-intervention, predicting optimal \ac{hp} configurations for installers and users, analyzing operational \ac{hp} flexibility, and enhancing load disaggregation methods.
By providing this dataset, we aim to support researchers and practitioners in developing innovative methods to optimize \ac{hp} performance in the field and create scalable digital services.

\begin{acks}
The research project is funded by the Swiss Federal Office of Energy under the grant number SI/502257. 
We thank Lorenz Deppeler, Hardy Schr\"oder, Michael Peter, and Nina Bichsel for their help on the data collection. 
Further, we acknowledge the support of other people involved: Jan Marckhoff, Tobias Graml, Dominik Bilgeri, Stephan Renz, Carina Alles, Rita Kobler, and Roland Br\"uniger.
\end{acks}

\subsection*{Code and Data Availability}
Available on GitHub: \href{https://github.com/tbrumue/heapo}{https://github.com/tbrumue/heapo}.

\subsection*{Author Contributions}
T.B.: conceptualization; data curation; formal analysis; investigation; methodology; software; visualization; writing - original draft; writing - review \& editing; project administration. 
E.F., M.V.G., \& T.S.: writing - review \& editing; supervision; project administration.

\balance
\bibliographystyle{ACM-Reference-Format}
\bibliography{bibliography}

\clearpage
\nobalance
\appendix
\section{Appendix}
\label{sec:appendix}

\subsection{Acronyms}
\begin{itemize}
	\item[] \acs{ami} \tabto{10em} \Acl{ami}
    \item[] \acs{ashp} \tabto{10em} \Acl{ashp}
    \item[] \acs{cdd} \tabto{10em} \Acl{cdd}
    \item[] \acs{cop} \tabto{10em} \Acl{cop}
    \item[] \acs{dhw} \tabto{10em} \Acl{dhw}
    \item[] \acs{dr} \tabto{10em} \Acl{dr}
    \item[] \acs{ekz} \tabto{10em} \Acl{ekz}
    \item[] \acs{eu} \tabto{10em} \Acl{eu}
    \item[] \acs{ev} \tabto{10em} \Acl{ev}
    \item[] \acs{ewh} \tabto{10em} \Acl{ewh}
    \item[] \acs{gshp} \tabto{10em} \Acl{gshp}
    \item[] \acs{hdd} \tabto{10em} \Acl{hdd}
    \item[] \acs{hp} \tabto{10em} \Acl{hp}
    \item[] \acs{iea} \tabto{10em} \Acl{iea}
    \item[] \acs{obis} \tabto{10em} \Acl{obis}
    \item[] \acs{pv} \tabto{10em} \Acl{pv}
    \item[] \acs{sfoe} \tabto{10em} \Acl{sfoe}
    \item[] \acs{sm} \tabto{10em} \Acl{sm}
    \item[] \acs{uk} \tabto{10em} \Acl{uk}
    \item[] \acs{us} \tabto{10em} \Acl{us}
    \item[] \acs{utc} \tabto{10em} \Acl{utc}
\end{itemize}

\subsection{Additional Explanations of \acs{obis} Codes}
\label{sec:appendix:obiscodes}
OBIS codes are standardized identification codes commonly used for measuring devices and data transmission, enabling the clear identification of measured values such as energy quantities or meter readings. 
This standard is widely adopted by utilities in German-speaking countries.  
An OBIS code consists of five value groups in the format \textbf{A-B:C.D.E}:  
\begin{itemize}
    \item \textbf{A} - Defines the medium (e.g., electricity; gas; water; heat).  
    \item \textbf{B} - Specifies internal or external channels.
    \item \textbf{C} - Indicates the measured variable (e.g., active, reactive, or apparent power; current; voltage).
    \item \textbf{D} - Represents the measurement type (e.g., maximum value; current value; time integral).
    \item \textbf{E} - Identifies the tariff category (e.g., total; tariff 1; tariff 2).
\end{itemize}
The \ac{smd} measurements in the \emph{HEAPO} dataset correspond to the following OBIS codes (excluding tariff-specific information):  

\begin{itemize}  
    \item \textbf{1-1:1.8.0} - Daily meter reading of active energy received from the grid (kWh).  
    \item \textbf{1-1:1.29.0} - 15-minute load profile of active energy received from the grid (kWh).
    \item \textbf{1-1:2.8.0} - Daily meter reading of active energy returned to the grid (kWh); available only for households with a \ac{pv} system and return delivery capability.  
    \item \textbf{1-1:5.8.0} - Daily meter reading of reactive inductive energy received from the grid (kvarh); refers to the first quadrant component.  
    \item \textbf{1-1:8.8.0} - Daily meter reading of reactive capacitive energy received from the grid (kvarh); refers to the fourth quadrant component.  
\end{itemize}  

For further details on \ac{obis} codes, see \cite{obiscode, 10.1007/978-981-10-8249-8_2}.

\subsection{Research Project}
\label{sec:appendix:dataprovider}
The public release of this dataset is part of the collaborative research project \emph{KIWO~--~K\"unstliche Intelligenz f\"ur die W\"armepumpenoptimierung (Artificial Intelligence for Heat Pump Optimization)}.  
Funded by the \ac{sfoe} under grant SI/502257\footnote{More details: \href{https://www.aramis.admin.ch/Grunddaten/?ProjectID=48862}{https://www.aramis.admin.ch/Grunddaten/?ProjectID=48862} (Last accessed: 2025 March 14).}, the project investigates the intersection of machine learning and \ac{smd} to provide households with personalized feedback on \ac{hp} electricity consumption and optimization strategies.  
The project is a collaboration between:  

\begin{enumerate}
\item \textbf{\href{https://www.bitstoenergy.com}{Bits to Energy Lab}}:  
   A research group leveraging technology for sustainability, with teams in Switzerland and Germany at ETH Zurich, the University of St. Gallen, the University of Bamberg, and FAU Erlangen-Nuremberg. The research group specializes in machine learning for energy applications and behavioral interventions to enhance user engagement.  

\item \textbf{\href{https://www.bfe.admin.ch/}{\acf{sfoe}}}:  
   Switzerland's leading authority on energy supply and policy, promoting renewable energy and efficiency. In 2024, the \ac{sfoe} planned CHF~480 million in investment and managed a CHF~1.3~billion grid surcharge fund.  

\item \textbf{\href{https://www.ekz.ch/}{\ac{ekz}}}:
   The utility for the canton of Zurich, supplying power to 125 municipalities directly and 35 through local distributors. With a 16,000 km grid, \ac{ekz} provides around 10\% of Switzerland's electricity to 1 million residents and businesses.  

\item \textbf{\href{https://www.enerlytica.com/}{Enerlytica (formerly BEN Energy)}}:  
   A leading data analytics company in the energy sector across Germany, Austria, and Switzerland, with over 150 projects and partnerships with 50+ companies.  

\item \textbf{\href{https://www.hoval.ch/}{Hoval}}:
   A global provider of heating and indoor climate solutions, specializing in \acp{hp}, solar systems, and multi-energy heating technologies.
\end{enumerate}

\subsection{Detailed Data Variable Descriptions}
The following tables provide detailed descriptions of the variables included in the metadata, protocol data, and weather data.


\begin{table}[hb]
\centering
\caption{Description of metadata variables.}
\label{table:metadatavariables}
\resizebox{\columnwidth}{!}{%
\begin{tabular}{@{}ll@{}}
\toprule
\textbf{Variable Name} & \textbf{Variable Description} \\ \midrule
Household\_ID & unique household identifier \\
Survey\_Building\_Type & type of building (house, appartment) \\
Survey\_Building\_LivingArea & living area of building ($m^2$) \\
Survey\_Building\_Residents & number of household residents \\
Survey\_HeatPump\_Installation\_Type & \acs{hp} type (air-source, ground-source) \\
Survey\_HeatDistribution\_System\_FloorHeating & heat distributed by floor heating \\
Survey\_HeatDistribution\_System\_Radiator & heat distributed by radiator \\
Survey\_DHW\_Production\_ByHeatPump & \acs{dhw} produced by \acs{hp} \\
Survey\_DHW\_Production\_ByElectricWaterHeater & \acs{dhw} produced by \acs{ewh} \\
Survey\_DHW\_Production\_BySolar & \acs{dhw} produced by solar collector \\
Survey\_Installation\_HasDryer & household has a dryer/tumbler \\
Survey\_Installation\_HasFreezer & household has a fridge-freezer \\
Survey\_Installation\_HasElectricVehicle & household has an \acs{ev} \\ \bottomrule
\end{tabular}%
}
\end{table}
\scriptsize
\onecolumn
\begin{longtable}[c]{@{}ll@{}}
\caption{Description of protocol data variables.}
\label{table:reportsvariablesdescription}\\
\toprule
\textbf{VariableName} & \textbf{Description} \\* \midrule
\endfirsthead
\multicolumn{2}{c}%
{{\bfseries Table \thetable\ continued from previous page}} \\
\endhead
\bottomrule
\endfoot
\endlastfoot
Report\_ID & unique report identifier \\
Household\_ID & unique household identifier \\
Visit\_Year & year of visit by energy consultant \\
Visit\_Date & date of visit by energy consultant \\
Building\_Type & type of building (single family house, multi family house, other) \\
Building\_HousingUnits & housing units of building \\
Building\_ConstructionYear & year of building construction \\
Building\_ConstructionYear\_Interval & interval of building construction year \\
Building\_Renovated\_Windows & indication if windows of building are renovated \\
Building\_Renovated\_Roof & indication if roof of building is renovated \\
Building\_Renovated\_Walls & indication if walls of building are renovated \\
Building\_Renovated\_Floor & indication if floor of building is renovated \\
Building\_FloorAreaHeated\_Total & total heated floor area of the building (in $m^2$) \\
Building\_FloorAreaHeated\_Basement & heated floor area of the basement of the building (in $m^2$) \\
Building\_FloorAreaHeated\_GroundFloor & heated floor area of the ground floor of the building (in $m^2$) \\
Building\_FloorAreaHeated\_FirstFloor & heated floor area of the first floor of the building (in $m^2$) \\
Building\_FloorAreaHeated\_SecondFloor & heated floor area of the second floor of the building (in $m^2$) \\
Building\_FloorAreaHeated\_TopFloor & heated floor area of the top floor of the building (in $m^2$) \\
Building\_FloorAreaHeated\_AdditionalAreasPlanned & indication if additional areas are planned to be heated any time soon \\
Building\_FloorAreaHeated\_AdditionalAreasPlannedSize & size of additional areas that are planned to be heated any time soon (in $m^2$) \\
Building\_Residents & number of household residents \\
Building\_PVSystem\_Available & indication if building has a \acs{pv} system \\
Building\_PVSystem\_Size & size of \acs{pv} system installed if available (in kWp) \\
Building\_ElectricVehicle\_Available & indication if household has one or more \acsp{ev} or plugin hybrid vehicles \\
HeatPump\_Installation\_Type & \acs{hp} type (air-source, ground-source, water-source) \\
HeatPump\_Installation\_Year & year of installation of \acs{hp} \\
HeatPump\_Installation\_Manufacturer & manufacturer of installed \acs{hp} \\
HeatPump\_Installation\_Model & model of installed \acs{hp} \\
HeatPump\_Installation\_HeatingCapacity & heating capacity of installed \acs{hp} (in kW) \\
HeatPump\_Installation\_Refrigerant\_Type & type of refrigerant used by \acs{hp} \\
HeatPump\_Installation\_Refrigerant\_Content & weight of refrigerant used by \acs{hp} (in kg) \\
HeatPump\_Installation\_Normpoint & operating point of \acs{hp} - according to product certificate \\
HeatPump\_Installation\_Normpoint\_COP & \acs{cop} of \acs{hp} at standardized operating point - according to product certificate \\
HeatPump\_Installation\_Normpoint\_ElectricPower & electric power of \acs{hp} at standardized operating point (in kW) - according to product certificate \\
HeatPump\_Installation\_Normpoint\_HeatingPower & thermal power of \acs{hp} at standardized operating point (in kW) - according to product certificate \\
HeatPump\_Installation\_Location & location of installed \acs{hp}(inside, outside, split) \\
HeatPump\_Installation\_InternetConnection & indication if \acs{hp} has a communication module, i.e., can deliver monitoring data via internet \\
HeatPump\_Installation\_ControllerNotAccessible & indication if \acs{hp} controller is not accessible due to password protection \\
HeatDistribution\_System\_Radiators & indication if heat is distributed by radiators \\
HeatDistribution\_System\_FloorHeating & indication if heat is distributed by floor heating \\
HeatDistribution\_System\_ThermostaticValve & indication if heat distribution uses thermostatic valves \\
HeatDistribution\_System\_BufferTankAvailable & indication if buffer tank is used for heat distribution \\
DHW\_Production\_ByHeatPump & indication if \acs{dhw} is produced by the \acs{hp} \\
DHW\_Production\_ByElectricWaterHeater & indication if \acs{dhw} is produced by \acs{ewh} \\
DHW\_Production\_BySolar & indication if \acs{dhw} is produced by solar collector \\
DHW\_Production\_ByHeatPumpBoiler & indication if \acs{dhw} is produced by \ac{hp} boiler \\
DHW\_ByHeatPump\_TimeInterval & time period that \acs{dhw} is responsible for \acs{dhw} production (whole year, half year, independent) \\
DHW\_Production\_TypeOfHeating & type of heating for \acs{dhw} production (central, decentral) \\
DHW\_Production\_Residents & amount of residents for which \acs{dhw} is produced \\
DHW\_Circulation\_NotInUse & indication if building does not use circulation for \acs{dhw} distribution \\
DHW\_Circulation\_ByTraceHeating & indication if circulation for \acs{dhw} distribution is done by trace heating \\
DHW\_Circulation\_ByCirculationPump & indication if circulation for \acs{dhw} distribution is done by circulation pump \\
DHW\_Circulation\_SwitchedByTimer & indication if circulation for \acs{dhw} distribution is switched by timer \\
DHW\_Sterilization\_Available & indication if sterilization system for \acs{dhw} production is available \\
DHW\_Sterilization\_Active & indication if sterilization system for \acs{dhw} production is not only available but also in use (i.e., active) \\
HeatPump\_Clean & indication if \acs{hp} is clean according to the impression of energy consultant during inspection \\
HeatPump\_BasicFunctionsOkay & indication if basic functionalities of \acs{hp} are okay or not as assessed by energy consultant during inspection \\
HeatPump\_TechnicallyOkay & indication if \acs{hp} is technically okay according to the impression of energy consultant during inspection \\
HeatPump\_ElectricityConsumption\_YearlyEstimated & estimation of yearly electricity consumption of \acs{hp} by energy consultant (in kWh) \\
HeatPump\_ElectricityConsumption\_Categorization & categorization of \acs{hp} energy consumption according to energy consultant (normal, rather high, rather low) \\
HeatPump\_Installation\_CorrectlyPlanned & indication if \acs{hp} is sized appropriately according to the energy consultant during inspection \\
HeatPump\_Installation\_IncorrectlyPlanned\_Categorization & categorization if \acs{hp} is oversized or undersized (if inappropriately sized according to the energy consultant) \\
HeatPump\_AirSource\_AirDuctsDistanceOkay & indication if distance between air ducts of \acs{ashp} is okay according to inspection by energy consultant \\
HeatPump\_AirSource\_AirDuctsFree & indication if air ducts of \acs{ashp} are free according to inspection by energy consultant \\
HeatPump\_AirSource\_AirDuctsCleaningRequired & indication if cleaning of air ducts of \acs{ashp} is necessary according to inspection by energy consultant \\
HeatPump\_AirSource\_AirDuctsDrainOkay & indication if drain of air ducts of \acs{ashp} is okay according to inspection by energy consultant \\
HeatPump\_AirSource\_EvaporatorClean & indication if evaporator of \acs{ashp} is clean according to inspection by energy consultant \\
HeatPump\_GroundSource\_BrineCircuit\_Length & length of brine circuit of \acs{gshp} \\
HeatPump\_GroundSource\_BrineCircuit\_Depth & depth of brine circuit of \acs{gshp} \\
HeatPump\_GroundSource\_BrineCircuit\_NumberOfHoles & number of drilled holes of brine circuit of \acs{gshp} \\
HeatPump\_GroundSource\_BrineCircuit\_CoolingCapacity & thermal capacity of ground probe of \acs{gshp} (in kW) \\
HeatPump\_GroundSource\_BrineCircuit\_AntiFreezeExists & indication if \acs{gshp} uses antifreeze \\
HeatPump\_GroundSource\_CurrentPressure & brine pressure of \acs{gshp} at time point of inspection by energy consultant \\
HeatPump\_GroundSource\_CurrentPressure\_Okay & indication if brine pressure of a \acs{gshp} is okay according to inspection by energy consultant \\
HeatPump\_GroundSource\_CurrentTemperature & brine temperature of \acs{gshp} at time point of inspection by energy consultant \\
HeatPump\_GroundSource\_CurrentTemperature\_Okay & indication if brine temperature of \acs{gshp} is okay according to inspection by energy consultant \\
HeatPump\_HeatingCurveSetting\_TooHigh\_BeforeVisit & indication if \acs{hp} heating curve setting was chosen too high before the visit of the energy consultant \\
HeatPump\_HeatingCurveSetting\_Changed & indication if energy consultant changed the heating curve setting (can be considered as reduction only) \\
HeatPump\_HeatingCurveSetting\_Outside20\_BeforeVisit & heating curve before visit of energy consultant: supply temperature (in \textdegree C) at outdoor temperature of 20 \textdegree C \\
HeatPump\_HeatingCurveSetting\_Outside0\_BeforeVisit & heating curve before visit of energy consultant: supply temperature (in \textdegree C) at outdoor temperature of 0 \textdegree C \\
HeatPump\_HeatingCurveSetting\_OutsideMinus8\_BeforeVisit & heating curve before visit of energy consultant: supply temperature (in \textdegree C) at outdoor temperature of -8 \textdegree C \\
HeatPump\_HeatingCurveSetting\_Outside20\_AfterVisit & heating curve after visit of energy consultant: supply temperature (in \textdegree C) at outdoor temperature of 20 \textdegree C \\
HeatPump\_HeatingCurveSetting\_Outside0\_AfterVisit & heating curve after visit of energy consultant: supply temperature (in \textdegree C) at outdoor temperature of 0 \textdegree C \\
HeatPump\_HeatingCurveSetting\_OutsideMinus8\_AfterVisit & heating curve after visit of energy consultant: supply temperature (in \textdegree C) at outdoor temperature of -8 \textdegree C \\
HeatPump\_HeatingLimitSetting\_TooHigh\_BeforeVisit & indication if \acs{hp} heating limit setting was chosen too high before the visit of the energy consultant \\
HeatPump\_HeatingLimitSetting\_Changed & indication if energy consultant changed the heating curve setting (can be considered as reduction only) \\
HeatPump\_HeatingLimitSetting\_BeforeVisit & setting of \acs{hp} heating limit before the visit of the energy consultant \\
HeatPump\_HeatingLimitSetting\_AfterVisit & setting of \acs{hp} heating limit after the visit of the energy consultant \\
HeatPump\_NightSetbackSetting\_Activated\_BeforeVisit & indication if \acs{hp} night setback was activated before visit(i.e., reduction of supply temperature at nighttime) \\
HeatPump\_NightSetbackSetting\_Activated\_AfterVisit & indication if \acs{hp} night setback is activated after visit (i.e., reduction of supply temperature at nighttime) \\
DHW\_TemperatureSetting\_Categorization & categorization of \acs{dhw} setpoint temperature value according to energy consultant (normal, too low, too high) \\
DHW\_TemperatureSetting\_Changed & indication if energy consultant changed setpoint temperature for \acs{dhw} during visit \\
DHW\_TemperatureSetting\_BeforeVisit & setpoint temperature for \acs{dhw} before visit of the energy consultant (in \textdegree C) \\
DHW\_TemperatureSetting\_AfterVisit & setpoint temperature for \acs{dhw} after visit of the energy consultant (in \textdegree C) \\
DHW\_Storage\_LastDescaling\_TooLongAgo & indication if storage system of \acs{dhw} was not descaled for a long time according to energy consultant \\
DHW\_Storage\_LastDescaling\_Year & year of last descaling of \acs{dhw} storage system \\
HeatDistribution\_ExpansionTank\_Pressure\_Categorization & categorization of preset pressure of expansion tank according to energy consultant (too high, too low, okay) \\
HeatDistribution\_ExpansionTank\_Pressure\_Actual & actual value of preset pressure of expansion tank (in bar) \\
HeatDistribution\_ExpansionTank\_Pressure\_Target & target value of expansion tank's preset pressure computed by energy consultant ((plant height + 3 meters) / 10) (in bar) \\
HeatDistribution\_ExpansionTank\_SystemHeight & plant height (used for calculating target value of preset pressure of expansion tank) (in meters) \\
HeatDistribution\_Circulation\_PumpStagePosition\_Changed & indication if energy consultant reduced the positional setting of the circulation pump for heat distribution \\
HeatDistribution\_Circulation\_PumpStagePosition\_BeforeVisit & positional setting of the circulation pump for heat distribution before visit of energy consultant \\
HeatDistribution\_Circulation\_PumpStagePosition\_AfterVisit & positional setting of the circulation pump for heat distribution after visit of energy consultant \\
HeatDistribution\_Recommendation\_InsulatePipes & indication if energy consultant recommends to better insulate pipes in the building \\
HeatDistribution\_Recommendation\_InstallThermostaticValve & indication if energy consultant recommends to install thermostatic valves for heat distribution \\
HeatDistribution\_Recommendation\_InstallRPMValve & indication if energy consultant recommends to install RPM-regulated circulation pump for heat distribution \\* \bottomrule
\end{longtable}
\normalsize
\begin{table*}[hb]
\centering
\caption{Description of weather data variables}
\label{table:weathervariables}
\resizebox{0.8\textwidth}{!}{%
\begin{tabular}{@{}llll@{}}
\toprule
\textbf{Variable Name} & \textbf{Parameter} & \textbf{Resolution} & \textbf{Variable Description} \\ \midrule
Weather\_ID &  &  & unique weather station identifier \\
Timestamp &  &  & UTC timestamp of measurement (\textit{YYYY-MM-DD HH:MM:SS}) \\ \midrule
Temperature\_avg\_hourly & \multirow{8}{*}{temperature [\textdegree C]} & hourly & average air temperature 2 m above ground \\
Temperature\_max\_daily &  & daily & maximum temperature 2 m above ground \\
Temperature\_min\_daily &  & daily & minimum air temperature 2 m above ground \\
Temperature\_avg\_daily &  & daily & average air temperature 2 m above ground \\
HeatingDegree\_SIA\_daily &  & daily & heating degree day number (Swiss standard SIA 381/3 \cite{articlea:standard1982381}) \\
HeatingDegree\_US\_daily &  & daily & heating degree day number (\acs{us} standard) \\
CoolingDegree\_US\_daily &  & daily & cooling degree day number (\acs{us} standard) \\
DewPoint\_hourly &  & hourly & Dew point 2 m above ground (instantaneous value) \\ \midrule
Humidity\_avg\_hourly & \multirow{2}{*}{humidity [\%]} & hourly & average relative humidity 2 m above ground \\
Humidity\_avg\_daily &  & daily & average relative humidity 2 m above ground \\ \midrule
Precipitation\_total\_hourly & \multirow{2}{*}{precipitation [mm]} & hourly & total precipitation \\
Precipitation\_total\_daily &  & daily & total precipitation \\ \midrule
Pressure\_BarometricHeight\_avg\_hourly & \multirow{3}{*}{air pressure [hPa]} & hourly & average air pressure at barometric height (QFE) \\
Pressure\_SeaLevelStandardAtmosphere\_avg\_hourly &  & hourly & average air pressure at sea level with standard atmosphere (QNH) \\
Pressure\_SeaLevel\_avg\_hourly &  & hourly & average air pressure at sea level (QFF) \\ \midrule
Sunshine\_duration\_hourly & \multirow{2}{*}{sunshine duration [h]} & hourly & total sunshine duration \\
Sunshine\_duration\_daily &  & daily & Sunshine duration - daily total \\ \midrule
WindSpeed\_hourly & wind [$m/s$] & hourly & average wind speed\\ \bottomrule
\end{tabular}%
}
\end{table*}

\newpage
\subsection{Folder and File Structure}
\begin{figure*}[ht!]
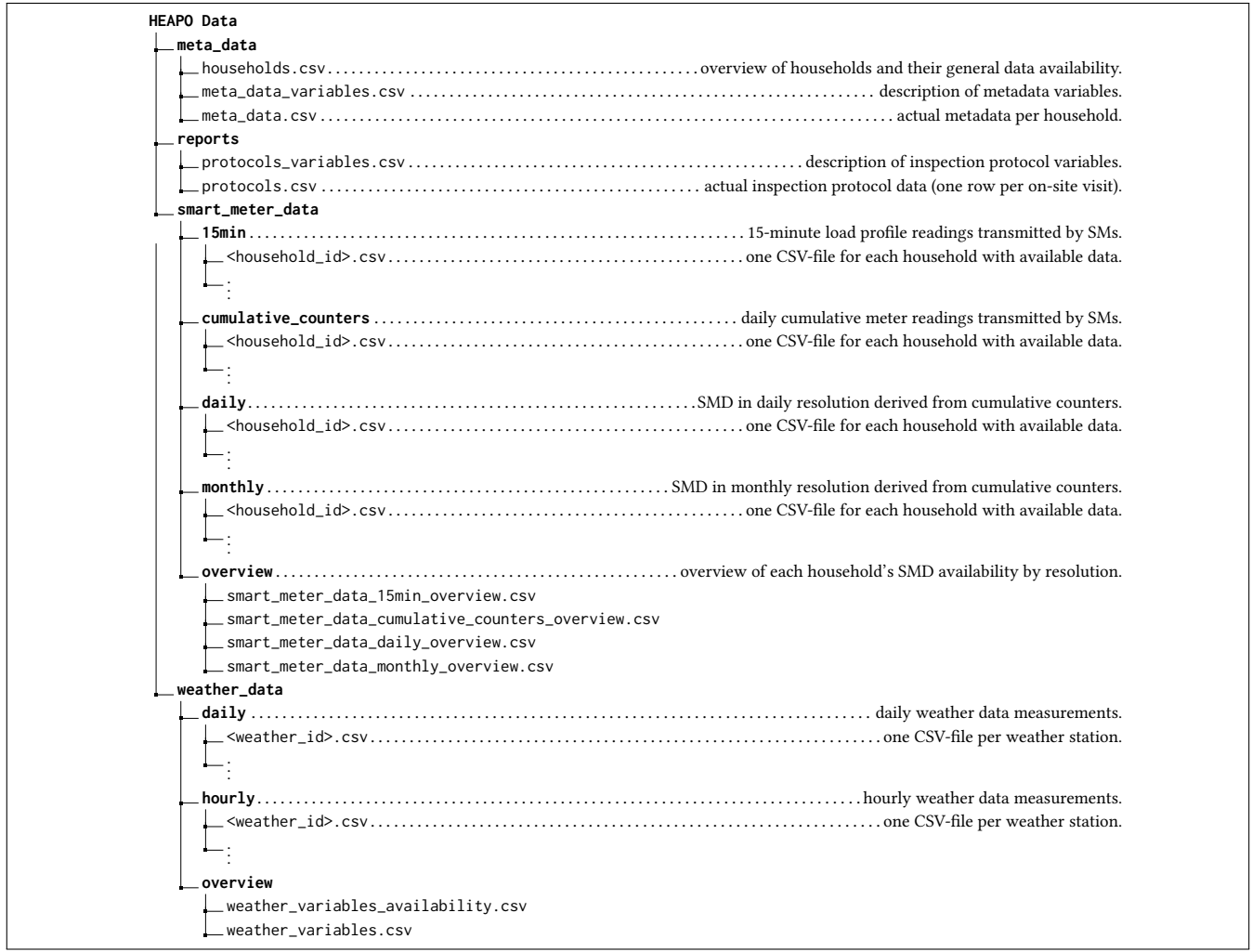

\centering
\framebox[1.0\textwidth]{%
\scalebox{0.8}{
\begin{minipage}{1.0\textwidth}
\dirtree{%
.1 \textbf{HEAPO Data}. 
.2 \textbf{meta\_data}. 
.3 \textit{households.csv}\DTcomment{overview of households and their general data availability.}. 
.3 \textit{meta\_data\_variables.csv}\DTcomment{description of metadata variables.}. 
.3 \textit{meta\_data.csv}\DTcomment{actual metadata per household.}. 
.2 \textbf{reports}.
.3 \textit{protocols\_variables.csv}\DTcomment{description of inspection protocol variables.}. 
.3 \textit{protocols.csv}\DTcomment{actual inspection protocol data (one row per on-site visit).}. 
.2 \textbf{smart\_meter\_data}.
.3 \textbf{15min}\DTcomment{15-minute load profile readings transmitted by \acp{sm}.}.
.4 \textit{<household\_id>.csv}\DTcomment{one CSV-file for each household with available data.}. 
.4 \textit{\vdots}. 
.3 \textbf{cumulative\_counters}\DTcomment{daily cumulative meter readings transmitted by \acp{sm}.}.
.4 \textit{<household\_id>.csv}\DTcomment{one CSV-file for each household with available data.}. 
.4 \textit{\vdots}. 
.3 \textbf{daily}\DTcomment{\ac{smd} in daily resolution derived from cumulative counters.}.
.4 \textit{<household\_id>.csv}\DTcomment{one CSV-file for each household with available data.}. 
.4 \textit{\vdots}. 
.3 \textbf{monthly}\DTcomment{\ac{smd} in monthly resolution derived from cumulative counters.}.
.4 \textit{<household\_id>.csv}\DTcomment{one CSV-file for each household with available data.}. 
.4 \textit{\vdots}. 
.3 \textbf{overview}\DTcomment{overview of each household's \ac{smd} availability by resolution.}.
.4 \textit{smart\_meter\_data\_15min\_overview.csv}.
.4 \textit{smart\_meter\_data\_cumulative\_counters\_overview.csv}.
.4 \textit{smart\_meter\_data\_daily\_overview.csv}.
.4 \textit{smart\_meter\_data\_monthly\_overview.csv}.
.2 \textbf{weather\_data}.
.3 \textbf{daily}\DTcomment{daily weather data measurements.}.
.4 \textit{<weather\_id>.csv}\DTcomment{one CSV-file per weather station.}. 
.4 \textit{\vdots}.
.3 \textbf{hourly}\DTcomment{hourly weather data measurements.}.
.4 \textit{<weather\_id>.csv}\DTcomment{one CSV-file per weather station.}. 
.4 \textit{\vdots}.
.3 \textbf{overview}.
.4 \textit{weather\_variables\_availability.csv}.
.4 \textit{weather\_variables.csv}.
}
\end{minipage}
}
}
\caption{File and folder structure of the \emph{HEAPO} dataset.}
\label{fig:filestructure}
\end{figure*}

Figure~\ref{fig:filestructure} presents the dataset's folder and file structure, as detailed in Section~\ref{sec:datasetstructure}.


\end{document}